\documentclass[prd,twocolumn]{revtex4}
\pdfoutput=1
\usepackage{graphicx, epsfig}
\usepackage{hyperref}
\usepackage{mathrsfs}
\usepackage{amsmath}

\newcommand{\be}{\begin{equation}}
\newcommand{\ee}{\end{equation}}
\newcommand{\bea}{\begin{eqnarray}}
\newcommand{\eea}{\end{eqnarray}}

\newcommand{\gapp}{\mathrel{\raise.3ex\hbox{$>$}\mkern-14mu
              \lower0.6ex\hbox{$\sim$}}}
\newcommand{\lapp}{\mathrel{\raise.3ex\hbox{$<$}\mkern-14mu
              \lower0.6ex\hbox{$\sim$}}}

\begin{document}
\title{Measuring the cosmological bulk flow using the peculiar velocities of supernovae}
\author{De-Chang Dai$^1$, William H. Kinney$^2$ and Dejan Stojkovic$^2$}
\affiliation{$^1$ Astrophysics, Cosmology and Gravity Centre, University of Cape Town, Rondebosch, Private Bag, 7700, South Africa}
\affiliation{$^2$ HEPCOS, Department of Physics,
SUNY at Buffalo, Buffalo, NY 14260-1500}


\begin{abstract}

We study large-scale coherent motion in our universe using the existing Type IA supernovae data. If the recently observed bulk flow is real, then some imprint must be left on supernovae motion. We perform a Bayesian Monte Carlo Markov Chain analysis in various redshift bins and find a sharp contrast between the $z < 0.05$ and $z > 0.05$ data. The $z < 0.05$ data are consistent with the bulk flow in the direction $(l,b)=({290^{+39}_{-31}}^{\circ},{20^{+32}_{-32}}^{\circ})$ with a magnitude of $v_{\rm bulk} = 188^{+119}_{-103}\ {\rm km/s}$ at $68\%$ confidence.  The significance of detection (compared to the null hypothesis) is  $95\%$. In contrast, $z > 0.05$ data (which contains $425$ of the $557$ supernovae in the Union2 data set) show no evidence for the bulk flow. While the direction of the bulk flow agrees very well with previous studies, the magnitude is significantly smaller. For example, the Kashlinsky, {\it et al.'s} original bulk flow result of $v_{\rm bulk} > 600\ {\rm km/s}$ is inconsistent with our analysis at greater than $99.7\%$ confidence level. Furthermore, our best-fit bulk flow velocity is consistent with the expectation for the $\Lambda$CDM model, which lies inside the $68\%$ confidence limit.
\end{abstract}


\maketitle

\section{Introduction}
One of the most recent puzzles raised by the cosmological observational data is the so-called ``bulk flow'' -- a coherent motion of a large part of our visible universe. Originally, Kashlinsky {\it et al.} \cite{Kashlinsky:2008ut,Kashlinsky:2009dw} claimed the detection of bulk flow using the Sunyaev-Zel'dovich effect in the Cosmic Microwave Background (CMB) radiation. The bulk flow velocity magnitude between $600\ {\rm km/s}$ and $1000\ {\rm km/s}$ directed toward the point given in galactic coordinates by $(l,b)=(283\pm 14^\circ, 12\pm 14^\circ )$ at scales surpassing $800$ Mpc was measured. This result was later reinforced in \cite{Watkins:2008hf,Feldman:2009es} by using a compilation of peculiar velocity redshift surveys. Their analysis showed a bulk flow of $407\pm 81\ {\rm km/s}$ towards $(l,b)=(287\pm 9^\circ, 8\pm 6^\circ )$, which is still large enough to be at odds with the $\Lambda$CDM model. Other work on the topic includes Refs. \cite{Hoffman:2001mb,Pike:2005tm,Strauss,RowanRobinson:1999mx,Lavaux,AtrioBarandela:2010wy,Kashlinsky:2010ur}. Though the directions of the bulk flow are very consistent, the magnitudes of the bulk flow velocities are less consistent and vary from analysis to analysis.

In this paper we study the peculiar velocities of Type IA supernovae. If the bulk flow is real, then some imprint must be left on supernovae motion. Since the recession velocities of the high redshift supernovae  are large and the magnitude of the bulk flow is moderate, one can expect the largest contribution to come from low redshift supernovae. Our results are consistent with this expectation. We find the bulk flow in the direction $(l,b)=({290^{+39}_{-31}}^{\circ},{20^{+32}_{-32}}^{\circ})$ with a magnitude of $v_{\rm bulk} = 188^{+119}_{-103}\ {\rm km/s}$ at $68\%$ confidence for the low redshift (i.e. $z < 0.05$) supernovae.

\section{Analysis}

The effects of Type IA supernovae peculiar velocities have been studied for a while \cite{Vanderveld:2008qu,Hui:2005nm} (for some earlier studies see \cite{Sasaki:1987ad,Pyne:1995ng}) because of their importance in cosmology as standard candles. In particular, the peculiar velocities can change the luminosity-distance relationship.
The measured (perturbed) luminosity-distance depends on the original (unperturbed) luminosity-distance and peculiar velocities of the observer and the source.
\begin{eqnarray}
\label{redshift}
z&=&\tilde{z}+(1+\tilde{z})\hat{n}\cdot (\vec{v}_s-\vec{v}_o),\\
\label{luminosity}
D_L(z)&=&(1+2\hat{n}\cdot \vec{v}_s-\hat{n}\cdot \vec{v}_o)\tilde{D}_L(\tilde{z}).
\end{eqnarray}
$\vec{v}_o$ and $\vec{v}_s$ are peculiar velocities of the observer and the source (supernova) respectively. $\hat{n}$ is the unit vector along the line of sight, and points from the observer to the supernova. $z$ and $D_L(z)$ are the redshift and luminosity distance from the Type Ia Supernova measurements. $\tilde{D}_L$ and $\tilde{z}$ are the unperturbed redshift and luminosity distance. Usually, the unperturbed frame is considered to be the CMB frame, which requires setting $\vec{v}_o =0$. One may also notice from Eq.~(\ref{luminosity}) that the physics is not invariant under the exchange of $\vec{v}_o$ and $\vec{v}_s$. This in principle gives the possibility to find out the locally preferred reference frame that moves with $\vec{v}_o$ with respect to the CMB frame.

 On large scales the universe can be described by an isotropic and homogeneous FLRW metric
\begin{equation}
ds^2=- dt^2 + a(t)^2 \left[dr^2 +r^2 (d\theta^2 +\sin^2 \theta d\phi^2)\right] .
\end{equation}
Here $a(t)$ is the scale factor, and we assume a flat metric. The supernovae luminosity distance relation in this metric can be written as

\begin{equation}
\label{lu_z}
D_L(z)=(1+z)\frac{c}{H_0}\int^{z}_0\frac{dz'}{\sqrt{\Omega_M(1+z')^3+\Omega_\Lambda}}
\end{equation}
where $c$ is the speed of light. The luminosity distance modulus is
\begin{equation}
m_B(z)=5\log_{10}\left[\frac{D_L(z)}{1{\rm Mpc}}\right]+25
\end{equation}

This is the relationship between the luminosity distance modulus, $m_B(z)$, and the redshift, $z$, that we will use in our analysis. This relationship depends on two velocity parameters $\vec{v}_o$ and $\vec{v}_s$. One of the ways to include the supernovae peculiar velocities is to have a different $\vec{v}_s$ for each individual supernova. However, we adopt here a more straightforward approach. We assume that for a large number of data points $\vec{v}_s$ can be treated as an average velocity of the whole system, in other words $\vec{v}_s$ is the bulk flow.

If the local reference frame coincides with the CMB frame (after we remove the CMB dipole), then $\vec{v}_o=0$. Then the only velocity parameter is $\vec{v}_s$, which is the bulk flow velocity. The bulk flow velocity as a vector is completely described by its magnitude and direction (in terms of galactic latitude and longitude). This implies that one would need to perform a three-parameter fit to the supernova data in order to find a possible bulk flow.  In a homogeneous universe, the cosmological parameters $\Omega_M$ and $\Omega_\Lambda $ are the same at all scales. The peculiar velocity may of course be anisotropic, but the ratio between the energy density components is not. However, it is still instructive to vary the values of the cosmological parameters to check for any possible degeneracies.  In principle, the uncertainties in the background parameters could result in effectively larger error bars on the luminosity distance, degrading the significance of the fit. We therefore perform a five-parameter fit in addition to the three-parameter fit, varying the values of cosmological parameters.
To obtain the level of significance of the $v_s \neq 0$ case versus the null hypothesis $v_s = 0$ we  calculate the one-parameter posterior probability on $v_s$, marginalized over all the other parameters in the fit.

\section{Results}

From the earlier work \cite{Sasaki:1987ad,Pyne:1995ng}, one can infer that the peculiar flow plays an important role for $z<0.1$. According to \cite{Kashlinsky:2008ut,Kashlinsky:2009dw} the bulk flow can extend to even higher redshifts up to($z<0.3$). Since the supernova data are readily available for these redshifts, we therefore expect the bulk flow to leak into the supernova measurements.

We analyze the Union2 data set from the Supernova Cosmology Project \cite{Amanullah:2010vv} using the Cosmomc \cite{Lewis:2002ah} code for implementing a Monte Carlo Markov chain search of the parameter space. We use a modified version of the likelihood code released with the Union2 data set, adding bulk flow parameters to the fit. We include full systematic errors for the data set, as described in \cite{Amanullah:2010vv}. 

\begin{figure}
\includegraphics[width=3.2in]{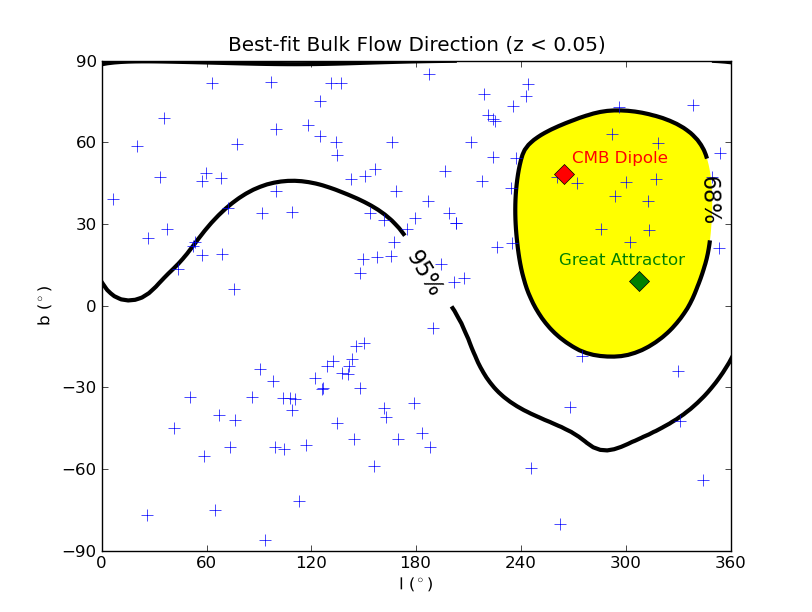}
\caption{Bulk flow direction in galactic longitude $l$ and galactic latitude $b$ for a three-parameter fit over $v_s$, $b$, and $l$, for redshifts $z < 0.05$.}
\label{bl3p}
\end{figure}

\begin{figure}
\includegraphics[width=3.2in]{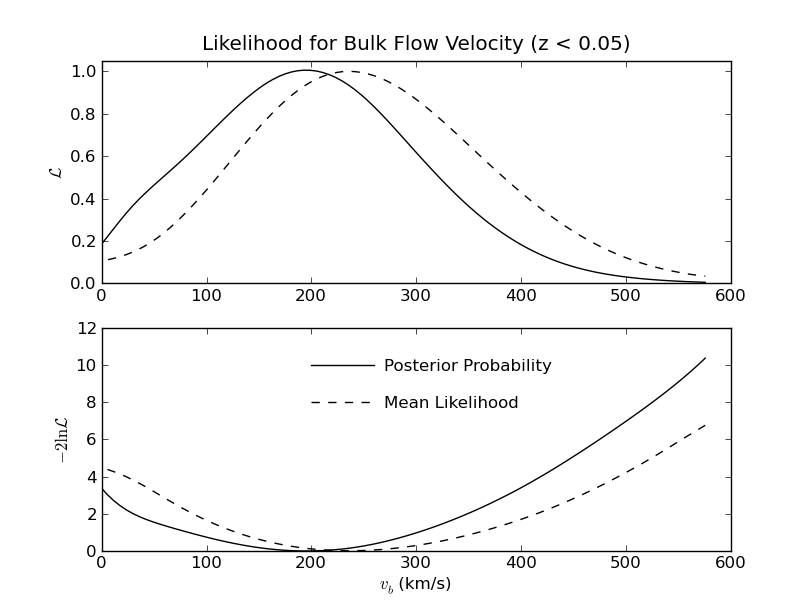}
\caption{Bulk flow magnitude $v_s$ for a three-parameter fit over $v_s$, $b$, and $l$, for redshifts $z < 0.05$.}
\label{vbulk3p}
\end{figure}

To perform the three-parameter fit to the data we fix the values of the cosmological parameters consistent with a best-fit from the WMAP 7-year data set \cite{Komatsu:2010fb,Larson:2010gs}, assuming $\Omega_M h^2 = 0.1334$, $w = -1$, and a flat universe. In both cases, we take $\Omega_b h^2 = 0.0226$, and make use of the HST measurement of the Hubble parameter $H_0 = 100h \text{ km s}^{-1} \text{Mpc}^{-1}$ \cite{hst} by multiplying the likelihood by a Gaussian likelihood function centered around $h=0.72$ and with a standard deviation $\sigma = 0.08$. We adopt flat priors on $l$ and $\sin(b)$. We divide the data into two sets, doing fits separately for $z < 0.05$ and for $z > 0.05$. Figures (\ref{bl3p}) and (\ref{vbulk3p}) show the best-fit direction and magnitude of the bulk flow for the three-parameter fit for $z < 0.05$.


\begin{figure}
\includegraphics[width=3.2in]{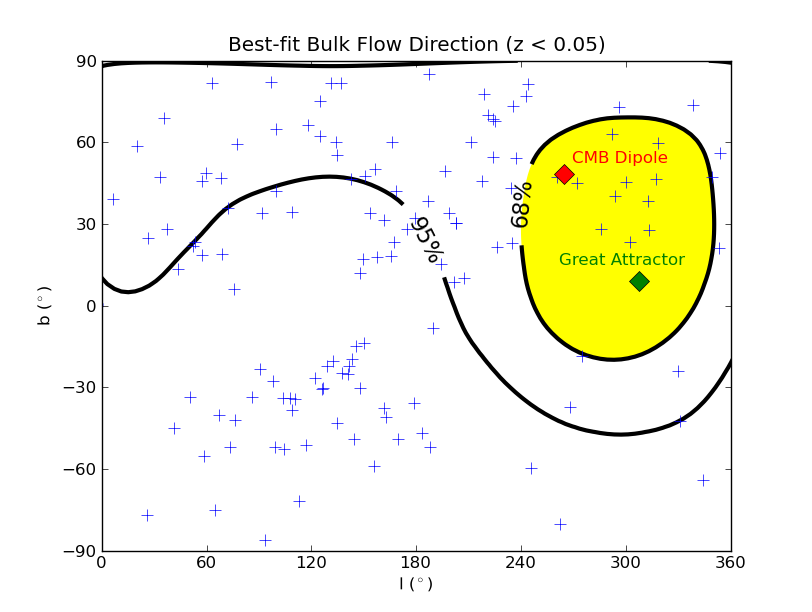}
\caption{Bulk flow direction in galactic longitude $l$ and galactic latitude $b$ for a five-parameter fit over $v_s$, $b$, $l$, $\Omega_M h^2$ and $w$, for redshifts $z < 0.05$. Crosses indicate the positions of the supernovae used in the fit, and the diamonds mark the directions of the CMB dipole and the Great Attractor.}
\label{bl5p}
\end{figure}

\begin{figure}
\includegraphics[width=3.2in]{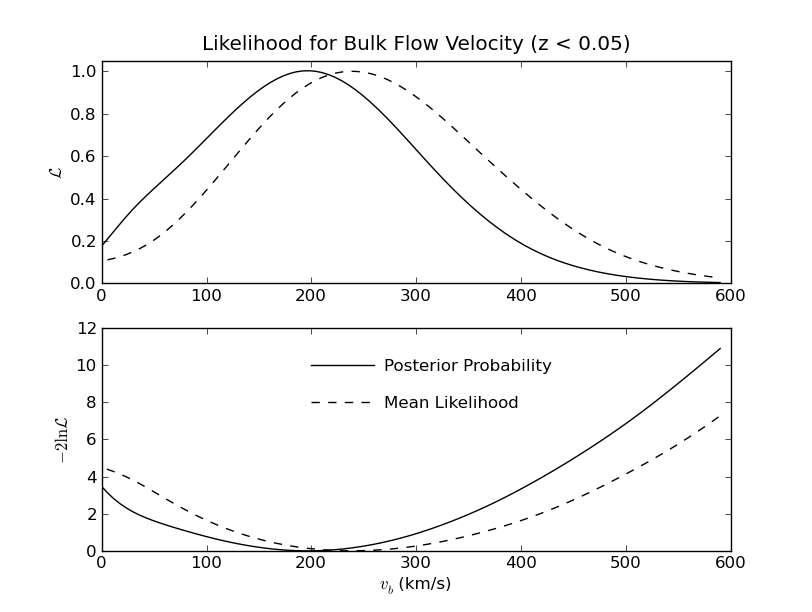}
\caption{Bulk flow magnitude $v_s$ for a five-parameter fit over $v_s$, $b$, $l$, $\Omega_M h^2$ and $w$, for redshifts $z < 0.05$.}
\label{vbulk5p}
\end{figure}

For the purpose of testing parameter degeneracies, we repeat the analysis assuming broad priors on $\Omega_M h^2$ and $w$
\begin{eqnarray}
0.01 < & \Omega_M h^2  & < 0.99 \\
-2 < & w & < -0.2.  \nonumber
\end{eqnarray}
We note that this is an unduly pessimistic assumption: inclusion of other data sets such as Baryon Acoustic Oscillations or CMB would constrain these parameters to much narrower regions. However, we are interested in investigating the effect of parameter degeneracies, and therefore allow the parameters to vary over a wide range. Figures (\ref{bl5p}) and (\ref{vbulk5p}) show the best-fit direction and magnitude of the bulk flow for the five-parameter fit with $z < 0.05$.
 
 
We first note, by comparing three- and five-parameter fits, that including $\Omega_M$ and $w$ does not add significant degeneracy to the system. This is reasonable
since it is unlikely that background parameters will be degenerate with parameters resulting in anisotropy. However, this is still an important check since the uncertainties in the background parameters could in principle degrade the significance of the fit.

\begin{figure}
\includegraphics[width=3.2in]{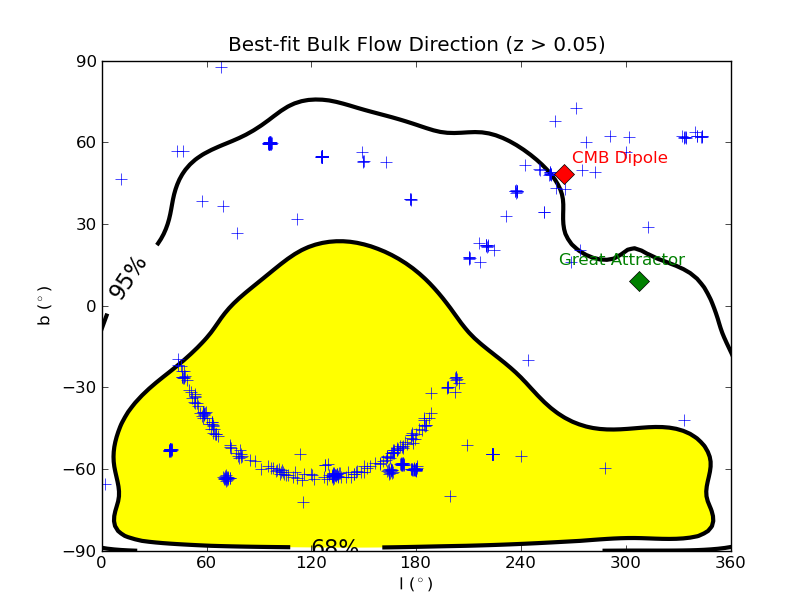}
\caption{Bulk flow direction in galactic longitude $l$ and galactic latitude $b$ for a three-parameter fit over $v_s$, $b$, and $l$, for redshifts $z > 0.05$.}
\label{bl3pzbig}
\end{figure}

\begin{figure}
\includegraphics[width=3.2in]{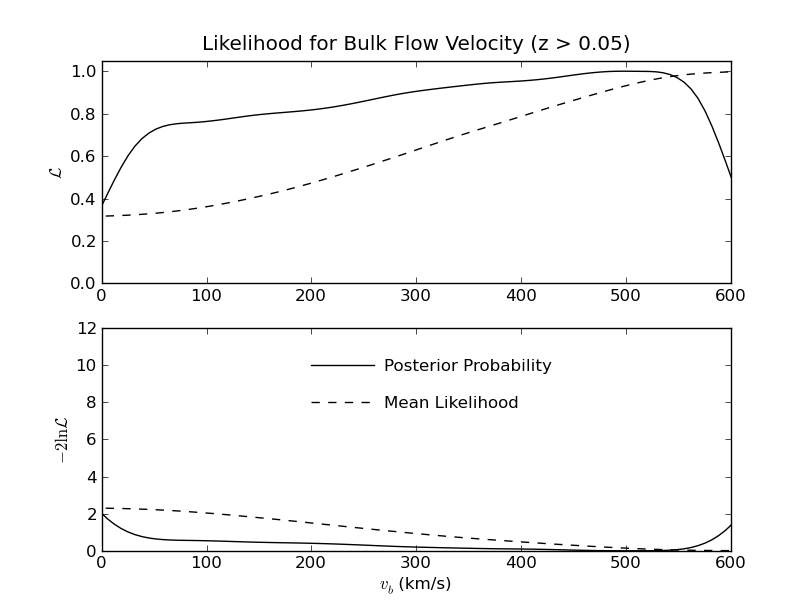}
\caption{Bulk flow magnitude $v_s$ for a three-parameter fit over $v_s$, $b$, and $l$, for redshifts $z > 0.05$.}
\label{vbulk3pzbig}
\end{figure}

The striking feature is a sharp contrast between the $z < 0.05$ and $z > 0.05$ data, shown in Figs. (\ref{bl3pzbig},\ref{vbulk3pzbig}). The $z < 0.05$ data are consistent with the bulk flow in the direction $(l,b)=({290^{+39}_{-31}}^{\circ},{20^{+32}_{-32}}^{\circ})$ with the magnitude of $v_{\rm bulk} = 188^{+119}_{-103}\ {\rm km/s}$ at $68\%$ confidence. In contrast, $z > 0.05$ data (which contains $425$ of the $557$ supernovae in the Union2 data set) show no evidence for the bulk flow. The significance of detection of $v_s \neq 0$ (compared to the null hypothesis $v_s = 0$) is  $95\%$ for $z < 0.05$, but gets progressively worse as we add higher redshift data. The absence of evidence for a bulk flow at higher redshifts may be a consequence of the fact that supernovae data is extremely sparse at large redshifts, and should not be viewed as the evidence of absence of the bulk flow at high redshifts. If our likelihood function in Fig.~\ref{vbulk3pzbig} were peaked at zero, we would be able to conclude that the data rule out bulk flow of a given velocity at a given confidence level. Instead, our likelihood function is almost flat, so the conservative conclusion is that we see neither evidence for bulk flow nor its absence. To settle this issue we certainly need more high redshift data.
The direction of the bulk flow for $z < 0.05$ agrees very well with previous studies \cite{Kashlinsky:2008ut,Kashlinsky:2009dw,Watkins:2008hf,Feldman:2009es}. However, the magnitude is significantly smaller. For example, Kashlinsky, {\it et al.}'s original bulk flow result of $v_{\rm bulk} > 600 \ {\rm km/s}$ is inconsistent with our analysis at greater than a $99.7\%$ confidence level. 
An interesting question is whether this much smaller magnitude is consistent with the expectation from a $\Lambda$CDM model. Our best-fit bulk flow velocity is around $200 \ {\rm km/s}$. 
The theoretical estimate of the rms amplitude of the peculiar velocity can be obtained from the power spectrum of primordial perturbations $P(k)$ as \cite{vrms}
\be
v_{\rm rms}= 4\pi H_0^2 f(\Omega)^2 \int P(k) W_v(kr) dk ,
\ee
where $f(\Omega)= 3 H_0 \Omega t$, while $W_v$ is the velocity window function usually taken to be a Fourier transform of the Gaussian window in $r$-space.
For the $\Lambda$CDM model, the rms amplitude of the peculiar velocity at some distance $d$ is \cite{Kashlinsky:2009dw}
\be
v_{\rm rms} = 250 (100 h^{-1}\ {\rm Mpc} / d) .
\ee
A redshift of $z = 0.05$ corresponds to $d=150 h^{-1}$ Mpc, so the expected  bulk flow velocity on that scale is about $170 \ {\rm
km/s}$, which is inside our $68\%$ confidence limit.
This implies that our findings are consistent with expectations from the $\Lambda$CDM model.

\section{Conclusions}

Recent results on cosmological peculiar velocities have garnered considerable attention. The original study of galaxy clusters using Sunyaev-Zel'dovich effect \cite{Kashlinsky:2008ut} and most of the follow up work report an unusually large bulk flow which is at odds with the $\Lambda$CDM model. There are however results questioning the original analysis \cite{Keisler:2009nw,Osborne:2010mf,Nusser:2011tu}. Any large scale coherent motion of matter in our universe must also be reflected in the peculiar motion of supernovae. In this paper we performed analysis of $557$ supernovae in the Union2 data set using the Cosmomc code, including full
systematic errors for the data set. For low redshift supernovae (i.e. $z < 0.05$) we do find a bulk flow in a direction consistent with previous work, however, with a much more moderate magnitude (we find $v_{\rm bulk} = 188^{+119}_{-103}\ {\rm km/s}$ at $68\%$ confidence, instead of the claimed $600-1000\ {
\rm km/s}$). Compared to the null hypothesis, the significance of detection is  $95\%$. We find no significant evidence for bulk flow for high redshift supernovae (i.e. $z > 0.05$).
This absence of evidence should not be understood as the evidence of absence. The current high redshift supernova data are too sparse to place a significant constraint on large-scale coherent motion. We also find that the expected bulk flow velocity of the $\Lambda$CDM model lies inside the $68\%$ confidence limit of our best-fit bulk flow velocity. Supernova data were also analyzed in this context in \cite{Colin:2010ds,Ma:2010ps,Haugboelle:2006uc,Davis:2010jq}, though using different data sets and/or different methods. The numerical results are mostly consistent with ours, though not necessarily the conclusions. We can summarize our results as follows: The $z < 0.05$ supernova data are completely consistent with a $\Lambda$CDM model and inconsistent with Kashlinsky, {\it et al.} \cite{Kashlinsky:2008ut} to high significance (greater than $99.7\%$ confidence level). The $z > 0.05$ data are inconclusive.

\begin{acknowledgments}
We thank Dragan Huterer for his valuable input. This research is supported in part by the National Science
Foundation under grants NSF-PHY-0757693 and NSF-PHY-0914893.
\end{acknowledgments}

\end{document}